\documentclass[aps,prd,amssymb,showpacs,floatfix,twocolumn,superscriptaddress,nofootinbib]{revtex4-2}
\usepackage{graphicx,amssymb,color}
\usepackage{multirow}

\newcommand{\black}{\protect\color{black}}
\newcommand{\red}{\black}

\begin{document}
\title{%
$K^{0}$ and $K^{+}$-meson electromagnetic form factors:
a nonperturbative relativistic quark model versus experimental,
perturbative and lattice Quantum-Chromodynamics results}
\author{S.V.~Troitsky}
\email{st@ms2.inr.ac.ru}
\affiliation{Institute
for Nuclear Research of the Russian Academy of Sciences, 60th October
Anniversary Prospect 7a, Moscow 117312, Russia}
\author{V.E.~Troitsky}
\email{troitsky@theory.sinp.msu.ru}
\affiliation{D.V.~Skobeltsyn
Institute of Nuclear Physics,\\
M. V. Lomonosov Moscow State University, Moscow 119991, Russia}
\begin{abstract}
It has been previously shown that a particular nonperturbative
constituent-quark model of hadrons describes
experimental measurements of electromagnetic form factors of light charged
mesons through a small number of common phenomenological parameters,
matching at the same time the Quantum-Chromodynamics (QCD) asymptotics for
the $\pi$-meson form factor at large momentum transfer. Here we start with
the determination of the $K^{0}$ electromagnetic form
factor in this approach. Precise measurement of the $K^{0}$ charge radius
makes it possible to constrain model parameters with high accuracy. Then,
with all parameters fixed, we revisit the $K^{+}$ form factor and find
that it matches experimental measurements in the infrared, lattice results
at moderate momentum transfer and the perturbative QCD asymptotics in the
ultraviolet. In this way we obtain a narrow constraint on the $K^{+}$
charge radius, $\langle r_{K^{+}}^{2} \rangle =
\red 0.403^{+0.007}_{-0.006}$~fm$^{2}$\black, and extend the successful
infrared-ultraviolet connection from $\pi$ to $K$ mesons.
\end{abstract}
\maketitle

\section{Introduction and outline}
\label{sec:intro}
Quantitative description of strongly coupled composite systems remains one
of the principal unsolved problems in particle physics. In Quantum
Chromodynamics (QCD), perturbative methods fail at low energies, where
this gauge theory of strong interactions becomes strongly coupled. The key
questions regarding the structure of hadrons and the origin of their
masses remain unanswered, see e.g.\ Ref.~\cite{Roberts2021}. Beyond QCD,
the same questions arise in numerous extensions of the Standard Model (SM)
of particle physics which predict composite particles bound by new strong
gauge forces. Testable first-principle quantitative predictions of these
models can only be obtained through lattice calculations, which are
however resource consuming. Most of detailed lattice calculations of
phenomenological importance are concentrated on the QCD case and are not
directly applicable to various beyond-SM theories. It is therefore
important to develop novel, ready to use quantitative approaches to the
description of strongly coupled particle bound states, even those
applicable to particular problems only. Predictions obtained within these
approaches can be tested on QCD, where experimental results are available.

These methods are however not abundant since they need to be fully
non-perturbative and have to link parameters of the fundamental theory,
defined at weak coupling (ultraviolet), to observables defined at strong
coupling (infrared). The present work aims to contribute to the
development of one of such methods.

The approach we follow makes use of a constituent-quark model which keeps
relativistic invariance at every step of calculations. This
nonperturbative method~\cite{KrTr-PRC2002, KrTr-PRC2003, EChAYa2009} is
based on Dirac Relativistic Hamiltonian Dynamics, see e.g.\
Ref.~\cite{KeisterPolyzou}, in its instant form. Relativistic invariance
is guaranteed by the use of modified impulse
approximation~\cite{KrTr-PRC2002}. The model was used to calculate the
electromagnetic form factor, $F_{\pi}$, of the $\pi$ meson as a function
of squared momentum transfer, $Q^{2}$, with a small number of parameters
as early as in 1998~\cite{KrTr-EurPhysJ2001}. Subsequent
measurements~\cite{exp-data} of $F_{\pi}(Q^{2})$ matched the model
predictions perfectly~\cite{KrTr-PRC2009}.

At large $Q^{2}$, when the constituent-quark masses are switched off,
$F_{\pi}(Q^{2})$ calculated in the model agrees with the perturbative QCD
prediction~\cite{Farrar, Efremov, Lepage}, both in the functional
form~\cite{KrTr-asymp} and numerically~\cite{PRD}. This is remarkable
because the correct ultraviolet asymptotics is reached automatically for
the same choice of parameters which describes infrared data.

Further studies within this approach included calculations of
electromagnetic properties of other light mesons \cite{rho, Kmeson2016},
as well as preliminary studies of the $\pi$-meson gravitational form
factor \cite{pi-gravi} and a generalization of the infrared-ultraviolet
$\pi$-meson link to gauge theories beyond QCD \cite{beyondQCD}, in a
general agreement with available lattice results. However, one of
parameters of the model for $K$ mesons remained poorly constrained because
of the low precision of the experimental determination of $K^{+}$ charge
radius,
$\langle r_{K^{+}}^{2} \rangle^{1/2}$.
In Ref.~\cite{Kmeson2016}, we presented the allowed range for the $K^{+}$
form factor,
$F_{K^{+}}(Q^{2})$,
and pointed out that more data are necessary to obtain firm predictions.
In particular, no conclusive study of the ultraviolet behaviour of
$F_{K^{+}}(Q^{2})$
was possible with that level of precision.

The present work fills this gap and makes use of much more precise
experimental measurements of the $K^{0}$ charge radius,
$\langle r_{K^{0}}^{2} \rangle^{1/2}$.
The calculation of the $K^{0}$ electromagnetic form factor is a
straightforward generalization of that for $K^{+}$, and
$F_{K^{0}}(Q^{2})$
is determined by the same set of parameters. In Sec.~\ref{sec:K0}, we use
the measurement of $\langle r_{K^{0}}^{2} \rangle^{1/2}$ to remove the
freedom in the only unconstrained parameter and make firm predictions for
$F_{K^{0}}(Q^{2})$. With all parameters fixed in this way, in
Sec.~\ref{sec:Kplus} we calculate and explore $F_{K^{+}}(Q^{2})$. Its
behaviour at $Q^{2} \to 0$ determines the $K^{+}$ charge radius, which we
constrain to a very narrow range. The precision is now sufficient to
demonstrate the agreement with recent lattice calculations of
$F_{K^{+}}(Q^{2})$ at $Q^{2} \sim$ a few GeV$^{2}$ and to determine
unambiguously the asymptotics $F_{K^{+}}(Q^{2}\to \infty)$, which we find
to agree precisely with the perturbative QCD prediction.
Section~\ref{sec:concl} summarizes main results of this work, stressing in
particular the importance of the obtained extension of the quantitative
infrared--ultraviolet connection from $\pi$ to $K$ mesons.

\section{$K^{0}$ form factor and parameter fixing}
\label{sec:K0}
General formalism and particular formulae for calculation of meson
electromagnetic form factors have been presented in previous
publications
\cite{KrTr-PRC2002, KrTr-EurPhysJ2001, KrTr-PRC2009, Kmeson2016,
BalandinaK},
see in particular Appendix~A of
Ref.~\cite{Kmeson2016} for the $K^{+}$ meson; expressions for $K^{0}$ are
obtained by a simple substitution of $u$ to $d$ quark charges and magnetic
moments. To calculate the electormagnetic form factor of a meson in our
approach, one has to fix the following parameters (see also
Table~\ref{tab:params} in Sec.~\ref{sec:concl}):

-- masses $M_{q_{1},q_{2}}$ of two constituent quarks $q_{1},q_{2}$
composing the meson;

-- wave-function confinement scale $b$;

-- values $\kappa_{q_{1},q_{2}}$ of anomalous magnetic moments of the two
quarks.

The values of $\kappa_{u,d,s}$ are determined a prior\red{i} \black from
model-independent sum rules~\cite{kappa}, see Ref.~\cite{Kmeson2016} for
more details, so they are not treated as free parameters. \red Here, we
use the most recent data from Ref.~\cite{PDG2020} to reevaluate
$\kappa_{s}$ and its uncertainties. \black Also, the value of the
light-quark constituent mass, $M_{u}=M_{d}=0.22$~GeV, was determined with
high accuracy in previous studies of the $\pi$-meson form factor (this
value was motivated also in Ref.~\cite{mass-spectrum} on different
grounds). \red Given the precision of measurement of $\pi$-meson related
quantities, their contribution to the error balance is expected to be
small compared to the $K$-meson ones. \black We are left with two
parameters for the $K$ mesons, $M_{s}$ and $b$. One combination of them is
determined from the requirement of reproducing the correct experimental
value of the meson decay constant $f_{K}$, cf.\
Refs.~\cite{Kmeson2016, f_K}, and the only remaining parameter is subject
to fit. Previous studies of the $\pi$ and $K^{+}$ mesons used lower-energy
measurements of the meson charge radius to fix it. While for the $\pi$
meson this was sufficient to determine the behaviour of $F_{\pi}(Q^{2})$
for all $Q^{2}$ and even to prove that the correct asymptotics is reached
at $Q^{2}\to 0$ if $M_{u}$ is switched off~\cite{PRD}, precision of
available measurements of $\langle r_{K^{+}}^{2} \rangle^{1/2}$ is too low
to make firm conclusions even at moderate $Q^{2}$. In
Ref.~\cite{Kmeson2016}, we determined the range of available values of
$F_{K^{+}}(Q^{2})$ and noted that future experiments, even at moderate
$Q^{2}$, will constrain the $K^{+}$ charge radius in this approach.

The confinement scale $b$ deserves some comments. First, its definition
depends on the particular choice of the meson wave function, but it has
been shown~\cite{KrTr-EurPhysJ2001} (see also Ref.~\cite{Schlumpf}) that
the result for the form factor does not depend on this choice, provided
the parameters are fixed in such a way that the correct decay constant is
reproduced. Here we use the same parametrisation of the wave function as
in Ref.~\cite{Kmeson2016}. Second, though the direct derivation of $b$ in
terms of underlying theory is not available, one expects that its value is
determined by masses and strong-interaction properties of the two
quarks composing the meson. Therefore, the values of $b$ are different for
$\pi$ and $K$ but are expected to be the same for $K^{+}$ and $K^{0}$.
This makes it possible to use $f_{K}$ and either $\langle r_{K^{0}}^{2}
\rangle$ or $\langle r_{K^{+}}^{2} \rangle$ to constrain $M_{s}$ and $b$
required for derivation of both $F_{K^{+}}(Q^{2})$ and $F_{K^{0}}(Q^{2})$.
Self-consistency of this approach will be checked with the asymptotics of
$F_{K^{+}}(Q^{2})$ in Sec.~\ref{sec:Kplus}.

\begin{figure}
\centering
\includegraphics[width=\columnwidth]{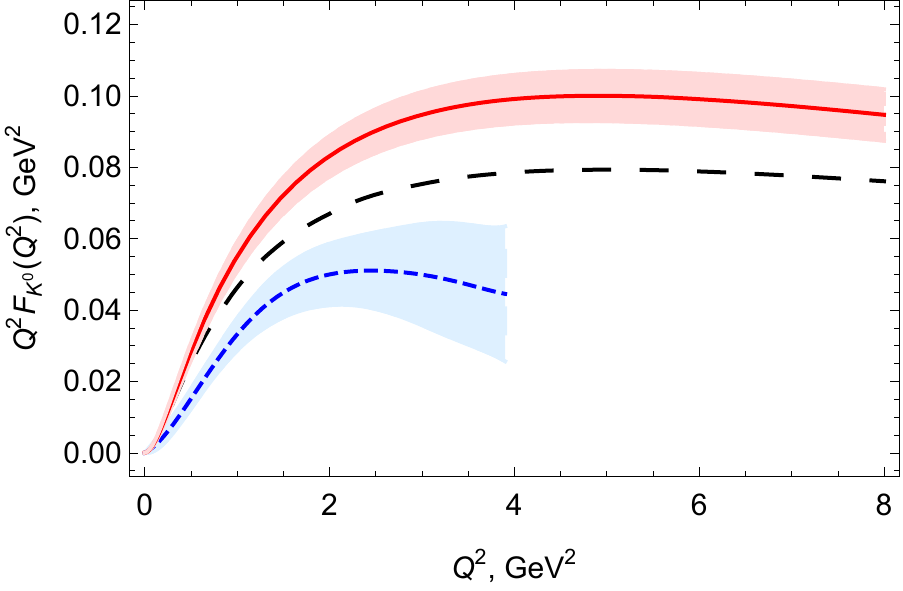}
\caption{
Results of the present work for the neutral kaon form factor (full red
line with light-red shadow uncertainty region) compared to the lattice
results \cite{Lattice2019} (dashed blue line with light-blue uncertainty
region) and
the calculation of Ref.~\cite{Roberts2017}  (long-da\red{sh}\black{}ed
black line). }
\label{fig:K0}
\end{figure}

We use the values of $f_{K}=0.1562\pm 0.0010$~GeV (same as in
Ref.~\cite{Kmeson2016} for consistency;
more recent estimates \cite{PDG2020} coincide with it within error bars)
and of
$\langle r_{K^{0}}^{2} \rangle=-0.077 \pm 0.010$~fm$^{2}$
\cite{PDG2020} to find $M_{s}=0.3177\pm 0.0040$~GeV and $b= 0.7118
\mp 0.0050$~GeV (error bars of $b$ and $M_{s}$ are correlated).
Figure~\ref{fig:K0}
presents $F_{K^{0}}(Q^{2})$ for these parameters in
comparison with the lattice result~\cite{Lattice2019} and with theoretical
calculations of Ref.~\cite{Roberts2017}.
A certain disagreement with the lattice result is explained by
the fact that the lattice calculation~\cite{Lattice2019} does not
reproduce the experimental value of the $K^{0}$ charge radius to which we
tuned our parameters: as it has been pointed out in
Ref.~\cite{Roberts2021}, the lattice value is
$\langle r_{K^{0}}^{2} \rangle \approx -0.026$~fm$^{2}$,
far beyond the experimental error bars.

\section{$K^{+}$ form factor from infrared to ultraviolet}
\label{sec:Kplus}
\begin{figure}
\centering
\includegraphics[width=\columnwidth]{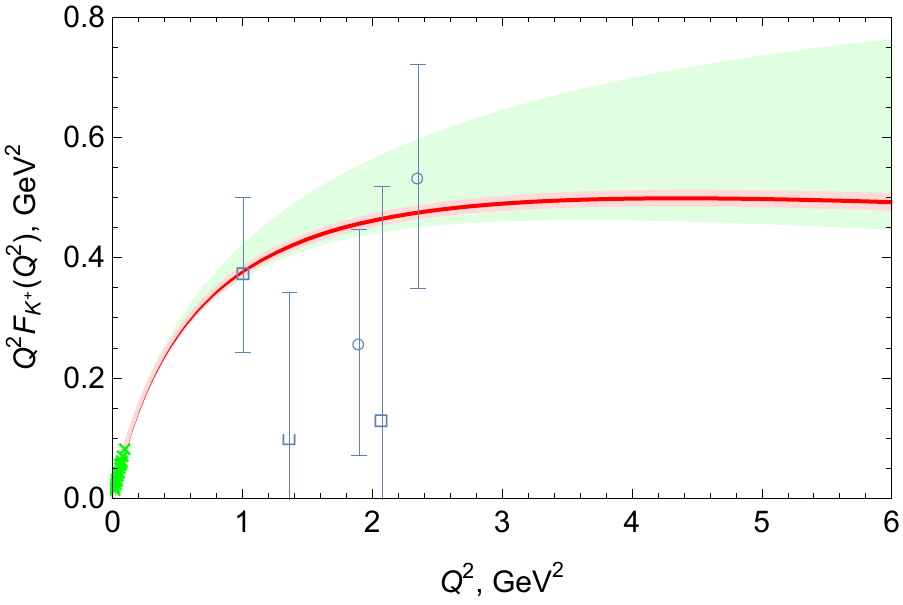}
\caption{
Results of the present work for the charged kaon form factor (full red
line with light-red shadow uncertainty region), with parameters fixed from
$K^{0}$ charge radius, compared to our previous
work~\cite{Kmeson2016} (light-green shading),
where the parameters were constrained from the $K^{+}$ charge radius.
\red Experimental measurements: \black green crosses at $Q^{2}\to 0$,
NA7 \cite{AmendoliaK}\red; squares, E93-018 \cite{Carmignotto}; circles,
E98-108 \cite{Carmignotto}\black. }
\label{fig:Klus-old-new}
\end{figure}

\begin{figure}
\centering
\includegraphics[width=\columnwidth]{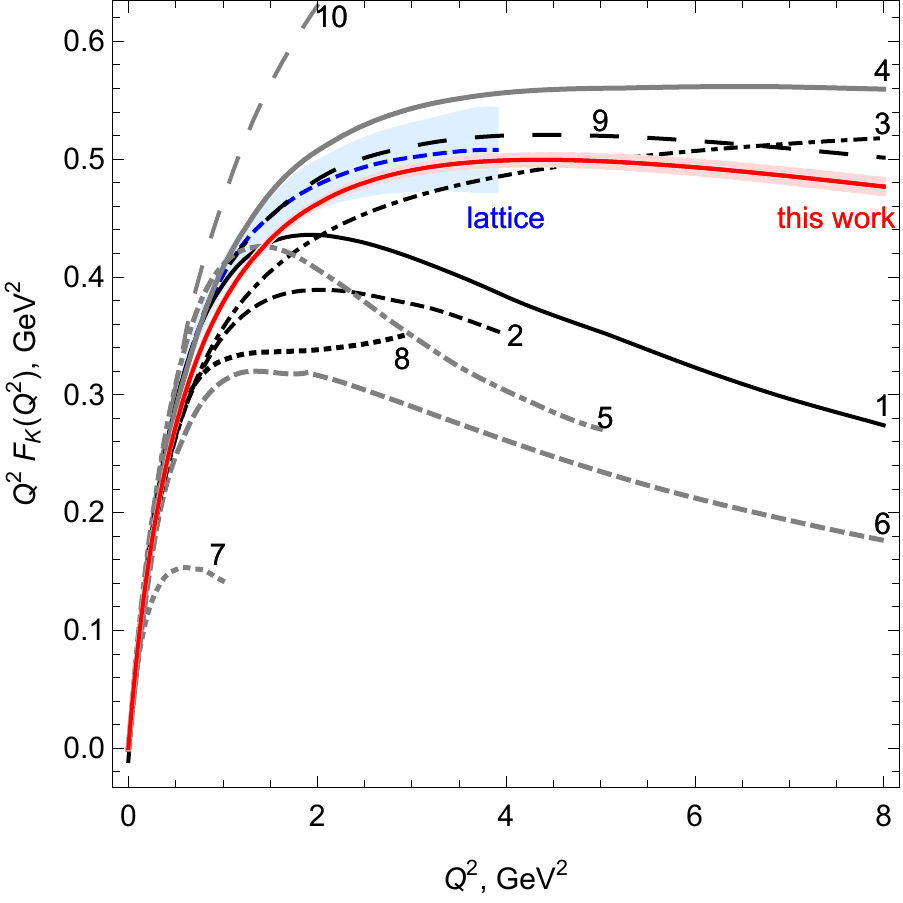}
\caption{
Results of the present work for the charged kaon form factor (full red
line with light-red shadow uncertainty region) compared to the lattice
results \cite{Lattice2019} (dashed blue line with light-blue uncertainty
region) and other theoretical studies --- 1 (full black): Ref.~\cite{1}; 2
(dashed black): Ref.~\cite{2}; 3 (dot-dashed black): Ref.~\cite{4}; 4
(full gray): Ref.~\cite{5}; 5 (dot-dashed gray): Ref.~\cite{6}; 6 (dashed
gray): Ref.~\cite{7}; 7 (dotted gray): Ref.~\cite{8}; 8 (dotted black):
Ref.~\cite{Galkin}; 9 (long-dashed black): Ref.~\cite{Roberts2017}; 10
(long-dashed gray): Ref.~\cite{AbidinHutauruk2019}. }
\label{fig:Kplus-comparison}
\end{figure}

With all parameters fixed in Sec.~\ref{sec:K0}, we proceed to calculate
the form factor of $K^{+}$. This repeats the calculation of
Ref.~\cite{Kmeson2016} but with all remaining freedom restricted to tiny
error bars in $M_{s}$ and $b$ specified in Sec.~\ref{sec:K0}. The resulted
$F_{K^{+}}(Q^{2})$
is compared to our previous work~\cite{Kmeson2016} in
Fig.~\ref{fig:Klus-old-new}.
The change in the precision is dramatic. Figure~\ref{fig:Kplus-comparison}
compares our result for
$F_{K^{+}}(Q^{2})$
with other theoretical calculations.

\begin{figure}
\centering
\includegraphics[width=\columnwidth]{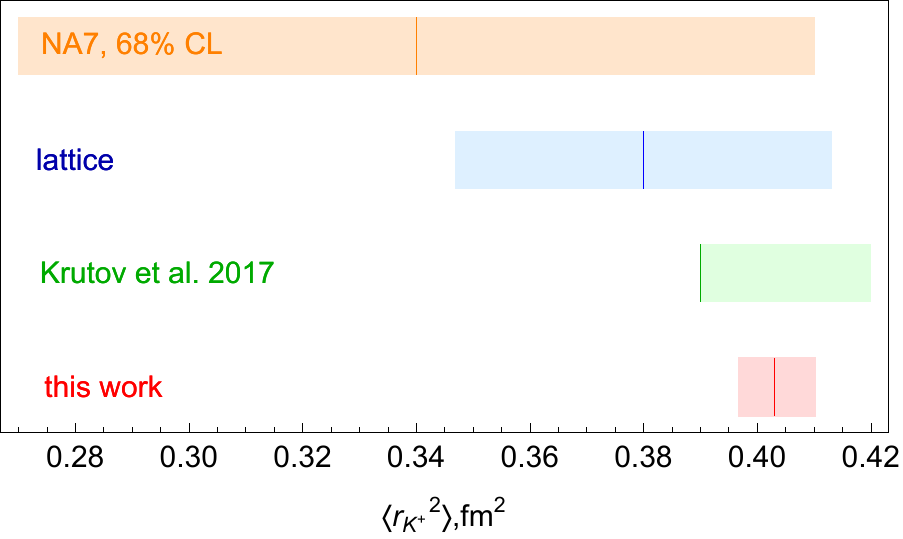}
\caption{
$K^{+}$ charge radius as determined by the NA7 experiment in
CERN \cite{AmendoliaK} (error bars scaled from reported 50\% CL to 68\% CL
assuming Gaussian distribution), by lattice calculations~\cite{lattice},
by a
fit of Ref.~\cite{AmendoliaK} data in our model (Krutov et al.\ 2017,
Ref.~\cite{Kmeson2016}) and by making use of constraints from $\langle
r_{K^{0}}^{2} \rangle$ in the present work.}
\label{fig:radius}
\end{figure}

Since the function
$F_{K^{+}}(Q^{2})$
is predicted without any freedom and with small uncertainties, it is
interesting to study its behaviour in more detail. First, we note that at
$Q^{2}\to 0$ it fits experimental measurements~\cite{AmendoliaK}. To be
more precise in this statement, we obtain the value of the charged kaon
charge radius predicted in our approach,
$\langle r_{K^{+}}^{2} \rangle =
\red 0.403^{+0.007}_{-0.006}$~fm$^{2}$\black    (68\%~CL).
Figure~\ref{fig:radius}
presents a graphical comparison of this value with the experimental
determination of
$\langle r_{K^{+}}^{2} \rangle$ \cite{AmendoliaK}, our previous
constraints~\cite{Kmeson2016}
and lattice results~\cite{lattice}. We note
that the value of $\langle r_{K^{+}}^{2} \rangle$ obtained in the present
work agrees well with all these previous results, while the precision is
improved considerably. It also agrees with less precise predictions of
chiral perturbation theory~\cite{Bijnens} and of a holographic
light-front model~\cite{1805.08911}.

Let us move now to moderate $Q^{2}$ of order a few GeV$^{2}$. Presently,
there are no precise experimental measurements of $F_{K^{+}}(Q^{2})$ in
this range, so we compare our result to those of the lattice QCD
calculation~\cite{Lattice2019}; Fig.~\ref{fig:Kplus-comparison}
demonstrates a good agreement (we will return to the lattice comparison in
Sec.~\ref{sec:concl}). Measurements of $F_{K^{+}}(Q^{2})$ at the
Jefferson Laboratory~\cite{JLABupgrade, HornRoberts2016} and the
Electron-Ion Collider \cite{EIC2021} will make it possible to test
our
predictions experimentally.

We turn now to the asymptotic regime, $Q^{2}\to\infty$.
\red
A way to merge the regime of relativistic constituent-quark models,
$Q^{2}\to 0$, and that of the perturbative QCD, $Q^{2}\to \infty$, is to
allow the quark mass to run from the constituent-quark mass $M$ in the
infrared to the current-quark mass $m \ll M$ in the ultraviolet, see e.g.\
Ref.~\cite{Kiss}. This implies switching \black
off constituent-quark masses in th\red{}e asymptotic \black regime. Within
our approach, switching the mass off resulted~\cite{KrTr-asymp} in the
correct functional form of $F_{\pi}(Q^{2} \to \infty)$ in agreement with
QCD predictions~\cite{QuarkCounting1, QuarkCounting2}. Moreover, we have
shown~\cite{PRD} that no matter how the masses are switched off, the
correct coefficient of the perturbative QCD asymptotics of
$F_{\pi}$~\cite{Farrar, Efremov, Lepage} is properly reproduced. Since all
parameters of the model are fixed in the low-$Q^{2}$ limit, this provides
a nontrivial infrared--ultraviolet link yet to be understood in a
fundamental theory. Extension of this approach to other gauge theories
beyond QCD~\cite{beyondQCD} resulted in a reasonable agreement with scarce
available lattice calculations. Here we make use of the precise
parameter-free prediction of $F_{K^{+}}(Q^{2})$ to test whether this
unusual result holds for the mesons other than $\pi$.

We follow the approach discussed in detail in Ref.~\cite{PRD}. We choose
the same family of functions $M(Q^{2})$, see Sec.~II.B of Ref.~\cite{PRD},
motivated by Refs.~\cite{Kiss, Doff:mass-switching}, which describe
switching constituent-quark masses off at high $Q^{2}$. The functional
shapes for $M_{u,d}(Q^{2})$ and $M_{s}(Q^{2})$ are the same, while their
values at $Q^{2}=0$ are fixed as described above.
\red
This implies in particular that $M(Q^{2}) \simeq M(0)$ for $Q^{2}\lesssim
8$~GeV$^{2}$, the range to be tested in coming experiments. \black We vary
two parameters of $M(Q^{2})$ shape, which describe where and how smoothly
the mass is turned off. In a complete analogy with what we observe for
$F_{\pi}(Q^{2})$ in Ref.~\cite{PRD}, cf.\ Fig.~4 of that work, we find
that $F_{K^{+}}(Q^{2}\to \infty)$ does not depend on these two parameters.
\red Note that the behaviour of $F_{K^{+}}(Q^{2})$ in the transition
region, before the asymptotics is reached, depends on the choice of the
$M(Q^{2})$ function. \black

We are now ready to answer the most intriguing question, whether our
approach reproduces the correct QCD asymtotics for $F_{K^{+}}(Q^{2})$ like
it does for $F_{\pi}(Q^{2})$. The asymptotics, determined in
Ref.~\cite{Lepage}, reads
\begin{equation}
F_{K^{+}}(Q^{2}\to \infty) \sim F_{K^{+}}^{\rm as} =
\frac{8\pi}{Q^{2}}f_{K}^{2} \alpha_{S}^{\rm{1-loop}} (Q^{2}),
\label{Eq:as}
\end{equation}
where $\alpha_{S}^{\rm{1-loop}} (Q^{2})$ is one-loop running
strong gauge coupling constant.

\begin{figure}
\centering
\includegraphics[width=\columnwidth]{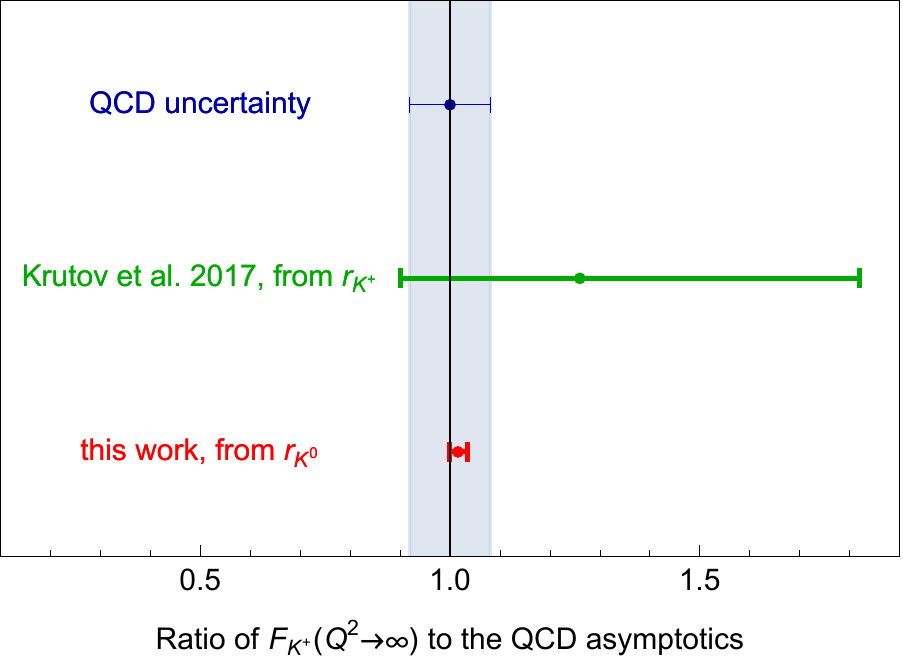}
\caption{
Ratio of the asymtotics of $F_{K^{+}}(Q^{2}\to \infty)$, obtained in our
model after switching off quark masses, to the perturbative QCD
asymptotics \cite{FarrarK}. Note that the correct functional form is
obtained automatically in our approach, and this plot presents the ratio
of the coefficients constrained in the previous study (Krutov et al.\
2017, Ref.~\cite{Kmeson2016}) and by making use of constraints from
$\langle r_{K^{0}}^{2} \rangle$ in the present work. See the text for the
estimate of the QCD uncertainty.}
\label{fig:asympt}
\end{figure}

The correct $Q^{2}$ dependence of the form factor is guaranteed, for
$M=0$, by analytical calculations of Ref.~\cite{KrTr-asymp}. We need to
calculate the coefficient numerically to compare it with that given by
Eq.~(\ref{Eq:as}). We start with the strong coupling at the $Z$-boson mass
scale, $\alpha_{S}(Q^2=M_{Z}^{2})= 0.1187 \pm 0.0018$~\cite{PDG2020}, and
evolve it down with the one-loop renormalization-group equation for QCD
with $N_{f}=5$ flavours. The uncertainty in Eq.~(\ref{Eq:as}) comes from a
combination of experimental error bars of $\alpha_{S}(M_{Z}^{2})$ and the
intrinsic theoretical uncertainty behind the one-loop
expression~(\ref{Eq:as}). The latter is not straightforward to estimate
since the generalization of Eq.~(\ref{Eq:as}) to the second loop is
nontrivial. To estimate the theoretical uncertainty in Eq.~(\ref{Eq:as}),
we loosely assume that it is of order of the ratio of the second
and first loop contributions to the beta function;
summed up in quadratures with the uncertainty
in $\alpha_{S}(M_{Z}^{2})$, this gives about 8\% estimated QCD uncertainty
at $Q_{0}^{2} \sim 200$~GeV$^{2}$. The ratio of $F_{K^{+}}(Q_{0}^{2})$,
calculated in our approach, to the asymptotics (\ref{Eq:as}) for
$\alpha_{S}(M_{Z}^{2})= 0.1187$ gives
\[
F_{K^{+}} / F_{K^{+}}^{\rm as} \simeq \red 1.016^{+0.020}_{-0.018}\black,
\]
well within the $1.00\pm 0.08$ QCD uncertainty (see Fig.~\ref{fig:asympt}).
This remarkable agreement extends our successful infra\-red--ultraviolet
link of $F_{\pi}$ \cite{PRD, beyondQCD} to the $K$-meson form factors.

\section{Discussion and conclusions}
\label{sec:concl}
We have calculated the neutral kaon electromagnetic form factor within the
same model and through the same parameters as it was previously done for
the charged kaon~\cite{Kmeson2016}. This has allowed us to fix the
remaining uncertain parameter of the model through the $K^{0}$ charge
radius, which is measured with much better precision than that of $K^{+}$.
As a result, we calculated also $F_{K^{+}}(Q^{2})$ with very small
uncertainty, see Figs.~\ref{fig:Klus-old-new}, \ref{fig:Kplus-comparison},
and without any remaining freedom in parameters, cf.\
Table~\ref{tab:params}.
\begin{table*}
\caption{\red Inputs and outputs of the present work. See
Ref.~\cite{Kmeson2016}
for details and the text for references concerning the fixed parameters.}
\label{tab:params}
\begin{tabular*}{\textwidth}{@{\extracolsep{\fill}}cccc@{}}
\hline
&&\\
&light-quark constituent mass&$M_{u}\equiv M_{d}=0.22$~GeV
                               & \multirow{3}{*}{$\left.\vphantom{
\begin{tabular}{c}1\\1\\
1\end{tabular}}\!\!\!\!\!\!\!\!\right\}$Fixed from previous
                               $\pi$-meson studies}\\
Fixed parameters& \multirow{3}{*}{quark anomalous magnetic moments
$\left\{\vphantom{\begin{tabular}{c}1\\1\\
1\end{tabular}}\!\!\!\!\!\!\!\!\right.$}& $\kappa_{u}=-0.01055$ &
\\
& &$\kappa_{\bar{d}}=0.03735$ & \\
& &$\kappa_{\bar{s}}=-0.084 \pm 0.019$ & Fixed from Ref.~\cite{kappa}
using data of Ref.~\cite{PDG2020}\\
 &&\\
\hline
&&\\
                 &  strange-quark constituent mass&$M_{s}=0.3177 \pm
0.0040$~GeV & \\
Fitted parameters&&                                          &
Fitted here from measured $f_{K}$ and $\langle r_{K^{0}}^{2} \rangle$\\
                 & $K$-meson wave-function scale &$b=0.7118\mp 0.0050$~GeV&
\\
&&&\\
\hline
&&&\\
&&\multicolumn{2}{c}{$Q^{2}\to 0$: refined $\langle r_{K^{+}}^{2} \rangle$, matches
measurements}\\
Outputs&Form factors $F_{K^{0}} (Q^{2})$, $F_{K^{+}} (Q^{2})$ &
\multicolumn{2}{c}{$Q^{2}\sim \mbox{few GeV}^{2}$:
matches lattice calculations}\\
&&\multicolumn{2}{c}{$Q^{2}\to \infty$: matches the perturbative QCD
asymptotics}\\
&&&\\
\hline
\end{tabular*}
\end{table*}
At $Q^{2}\to 0$, this calculation matches experimental and lattice results
for the $K^{+}$ charge radius, at the same time predicting its value with
much higher precision, cf.\ Fig.~\ref{fig:radius}. \red The values of the
form factors calculated in this work are given in Table~\ref{tab:values}
in Appendix~\ref{sec:app} together with their uncertainties. \black

At moderate $Q^{2} \lesssim 4$~GeV$^{2}$, we compare our results for
$F_{K^{0},K^{+}}(Q^{2})$ with lattice QCD calculations of
Ref.~\cite{Lattice2019}, see Figs.~\ref{fig:K0},
\ref{fig:Kplus-comparison}. Our $F_{K^{+}} (Q^{2})$ agrees very well with
the lattice result, while the agreement is much worse for $F_{K^{0}}
(Q^{2})$. This is not surprising since the lattice result does not
reproduce the measured $K^{0}$ charge radius, which we used to fit the key
parameter of our model. In a deeper context, this disagreement may be
related to the fact that the lattice $K^{0}$ form factor was obtained as a
difference between contributions from two quarks, weighted by their
charges. We note that in our approach, electromagnetic properties of the
quarks differ not only by their charges but also by their anomalous
magnetic moments; form factors of $K^{0}$ and $K^{+}$ are therefore not
decomposed into charge-weighted linear combinations of contributions from
each quark. \red The contribution of the anomalous magnetic moments is
$\sim 6\%$ for $K^{+}$ while for $K^{0}$, it exceeds $\sim 50\%$. \black
This explains in particular the fact that $F_{K^{+}}(Q^{2})$ agrees with
the lattice results while $F_{K^{0}}(Q^{2})$ does not agree, which is \red
also \black the case for the results of Refs.~\cite{Roberts2021,
Roberts2017}.

At $Q^{2}\to \infty$, upon switching constituent-quark masses off,
$F_{K^{+}}(Q^{2})$ reaches precisely the perturbative QCD asymptotics,
cf.\ Fig.~\ref{fig:asympt}. This is achieved without any parameter tuning:
all free parameters of the model are fixed in the infrared. This is
similar to the behaviour of $F_{\pi}(Q^{2})$ found previously in our
approach: the asymptotics are different but both are achieved
automatically. Our calculation does not predict the scale of the momentum
transfer at which the transition to the perturbative regime happens: it is
related to the scale at which $M(Q^{2})$ goes to zero; but the correct
asymptotics is reached independently of that scale. This scale, predicted
to be as low as $Q^{2}\simeq 3.5$~GeV$^{2}$ for
$F_{\pi}$~\cite{JacobKiss}, is probably actually higher, as it is
suggested by measurements of $F_{\pi}(Q^{2})$~\cite{exp-data} and
lattice calculations of $F_{K^{+}}(Q^{2})$~\cite{Lattice2019}. To
determine this scale in our model, further experimental input is
required~\cite{Troitsky:transition}. However, the relations between
infrared and ultraviolet properties of the form factors are valid
independently of their behaviour at moderate $Q^{2}$. This makes it
possible to apply these relations in new-physics models with
strongly-coupled gauge theories~\cite{beyondQCD}. In particular, these
include models with composite Higgs and dark-matter particles, see e.g.\
Refs.~\cite{compDMH, compDM-F}: there, the low-$Q^{2}$ form factor of a
neutral meson-like state, similar to $K^{0}$, determines observable
properties relevant for the dark-matter search. This avenue will be
explored elsewhere.

\appendix
\section{Numerical results for the $K^{0}$ and $K^{+}$ form factors}
\label{sec:app}
Table~\ref{tab:values}
\begin{table}
\caption{\red Values of $K^{0}$ and $K^{+}$ electromagnetic form factors
obtained in this work. }
\label{tab:values}
\begin{tabular*}{\columnwidth}{@{\extracolsep{\fill}}ccc@{}}
\hline
 $Q^{2}$, GeV$^{2}$ & $F_{K^{0}}$ & $F_{K^{+}}$  \\
\hline
0.0  &   0.0000 $\pm$0.0000  &1.0000  $+$0.0000 $-$0.0000 \\
0.1  &   0.0262 $\pm$0.0029  &0.8509  $+$0.0006 $-$0.0006 \\
0.2  &   0.0412 $\pm$0.0041  &0.7444  $+$0.0009 $-$0.0009 \\
0.3  &   0.0497 $\pm$0.0046  &0.6628  $+$0.0012 $-$0.0012 \\
0.4  &   0.0544 $\pm$0.0048  &0.5978  $+$0.0014 $-$0.0013 \\
0.5  &   0.0567 $\pm$0.0047  &0.5447  $+$0.0015 $-$0.0015 \\
0.6  &   0.0576 $\pm$0.0047  &0.5003  $+$0.0016 $-$0.0015 \\
0.7  &   0.0576 $\pm$0.0045  &0.4626  $+$0.0017 $-$0.0016 \\
0.8  &   0.0571 $\pm$0.0044  &0.4302  $+$0.0018 $-$0.0017 \\
0.9  &   0.0562 $\pm$0.0042  &0.4019  $+$0.0018 $-$0.0017 \\
1.0  &   0.0550 $\pm$0.0041  &0.3770  $+$0.0019 $-$0.0018 \\
1.5  &   0.0481 $\pm$0.0034  &0.2871  $+$0.0021 $-$0.0019 \\
2.0  &   0.0416 $\pm$0.0028  &0.2305  $+$0.0021 $-$0.0020 \\
2.5  &   0.0361 $\pm$0.0024  &0.1916  $+$0.0022 $-$0.0020 \\
3.0  &   0.0316 $\pm$0.0021  &0.1631  $+$0.0022 $-$0.0020 \\
3.5  &   0.0279 $\pm$0.0019  &0.1415  $+$0.0022 $-$0.0020 \\
4.0  &   0.0248 $\pm$0.0017  &0.1244  $+$0.0021 $-$0.0020 \\
4.5  &   0.0222 $\pm$0.0015  &0.1107  $+$0.0021 $-$0.0019 \\
5.0  &   0.0200 $\pm$0.0014  &0.0994  $+$0.0021 $-$0.0019 \\
5.5  &   0.0181 $\pm$0.0013  &0.0900  $+$0.0020 $-$0.0018 \\
6.0  &   0.0165 $\pm$0.0012  &0.0820  $+$0.0020 $-$0.0018 \\
6.5  &   0.0151 $\pm$0.0011  &0.0751  $+$0.0019 $-$0.0017 \\
7.0  &   0.0139 $\pm$0.0010  &0.0692  $+$0.0017 $-$0.0016 \\
7.5  &   0.0128 $\pm$0.0010  &0.0640  $+$0.0017 $-$0.0015 \\
8.0  &   0.0118 $\pm$0.0009  &0.0594  $+$0.0016 $-$0.0014 \\
\hline
\end{tabular*}
\end{table}
presents the form factors $F_{K^{0}}(Q^{2})$, $F_{K^{+}}(Q^{2})$,
determined in this work, in the numerical form.

\begin{acknowledgements}
This research has been supported by the Interdisciplinary
Scientific and Educational School of Moscow University ``Fundamental and
Applied Space Research''.
We thank A.~Krutov for fruitful discussions. ST thanks A.~Kataev
for bringing $K^{0}$ results to his attention.
\end{acknowledgements}


\begin{thebibliography}{51}

\bibitem{Roberts2021}
C.~D.~Roberts, D.~G.~Richards, T.~Horn and L.~Chang,
``Insights into the Emergence of Mass from Studies of Pion and Kaon
Structure,'' [arXiv:2102.01765 [hep-ph]].

\bibitem{KrTr-PRC2002}
A.~F.~Krutov and V.~E.~Troitsky,
``Relativistic instant form approach to the structure of two-body composite systems,''
  Phys.\ Rev.\ C {\bf 65} (2002) 045501
  [hep-ph/0204053].

\bibitem{KrTr-PRC2003}
A.~F.~Krutov and V.~E.~Troitsky,
``Relativistic instant form approach to the structure of two-body composite systems. 2. Nonzero spin,''
  Phys.\ Rev.\ C {\bf 68} (2003) 018501
  [hep-ph/0210046].

\bibitem{EChAYa2009}
 A.~F.~Krutov and V.~E.~Troitsky,
``Instant form of Poincare-invariant quantum mechanics and description of the structure of composite systems,''
  Phys.\ Part.\ Nucl.\  {\bf 40} (2009) 136.

\bibitem{KeisterPolyzou}
  B.~D.~Keister and W.~N.~Polyzou,
``Relativistic Hamiltonian dynamics in nuclear and particle physics,''
  Adv.\ Nucl.\ Phys.\  {\bf 20} (1991) 225.

\bibitem{KrTr-EurPhysJ2001}
A.~F.~Krutov and V.~E.~Troitsky,
 ``On a possible estimation of the constituent quark parameters from
  Jefferson Lab experiments on pion form-factor,'' Eur.\ Phys.\ J.\ C {\bf
  20} (2001) 71 [hep-ph/9811318].

\bibitem{exp-data}
G.~M.~Huber {\it et al.}  [Jefferson Lab Collaboration],
  ``Charged pion form-factor between $Q^2 = 0.60$~GeV$^{2}$ and
  2.45~GeV$^{2}$. II. Determination of, and results for, the pion
  form-factor,'' Phys.\ Rev.\ C {\bf 78} (2008) 045203 [arXiv:0809.3052
  [nucl-ex]].

\bibitem{KrTr-PRC2009}
A.~F.~Krutov, V.~E.~Troitsky and N.~A.~Tsirova,
``Nonperturbative relativistic approach to pion form factor versus JLab experiments,''
  Phys.\ Rev.\ C {\bf 80} (2009) 055210
  [arXiv:0910.3604 [nucl-th]].

\bibitem{Farrar}
G.~R.~Farrar and D.~R.~Jackson,
``The Pion Form-Factor,''
Phys. Rev. Lett. \textbf{43} (1979) 246

\bibitem{Efremov}
A.~V.~Efremov and A.~V.~Radyushkin,
``Factorization and Asymptotical Behavior of Pion Form-Factor in QCD,''
Phys. Lett. B \textbf{94} (1980) 245

\bibitem{Lepage}
G.~P.~Lepage and S.~J.~Brodsky,
``Exclusive Processes in Quantum Chromodynamics: Evolution Equations for Hadronic Wave Functions and the Form-Factors of Mesons,''
Phys. Lett. B \textbf{87} (1979) 359

\bibitem{KrTr-asymp}
A.~F.~Krutov and V.~E.~Troitsky,
``Asymptotic estimates of the pion charge form-factor,''
  Theor.\ Math.\ Phys.\  {\bf 116} (1998) 907
   [Teor.\ Mat.\ Fiz.\  {\bf 116} (1998) 215].

\bibitem{PRD}
S.~V.~Troitsky and V.~E.~Troitsky,
``Transition from a relativistic constituent-quark model to the
quantum-chromodynamical asymptotics: a quantitative description of the
pion electromagnetic form factor at intermediate values of the momentum
transfer,''
  Phys.\ Rev.\ D {\bf 88} (2013) 093005
  [arXiv:1310.1770 [hep-ph]].

\bibitem{rho}
 A.~F.~Krutov, R.~G.~Polezhaev and V.~E.~Troitsky,
``The radius of the $\rho$ meson determined from its decay constant,''
  Phys.\ Rev.\ D {\bf 93} (2016)   036007
  [arXiv:1602.00907 [hep-ph]].

\bibitem{Kmeson2016}
A.~F.~Krutov, S.~V.~Troitsky and V.~E.~Troitsky,
``The $K$-meson form factor and charge radius: linking low-energy data to future Jefferson Laboratory measurements,''
Eur. Phys. J. C \textbf{77} (2017) 464
[arXiv:1610.06405 [hep-ph]].

\bibitem{pi-gravi}
A.~F.~Krutov and V.~E.~Troitsky,
``Pion gravitational form factors in a relativistic theory of composite particles,''
Phys. Rev. D \textbf{103} (2021) 014029
[arXiv:2010.11640 [hep-ph]].

\bibitem{beyondQCD}
S.~Troitsky and V.~Troitsky,
``Linking infrared and ultraviolet parameters of pion-like states in strongly coupled gauge theories,''
Eur. Phys. J. C \textbf{78} (2018) 899
[arXiv:1807.11009 [hep-ph]].

\bibitem{BalandinaK}
E.~V.~Balandina, A.~F.~Krutov and V.~E.~Troitsky,
``Elastic charge form-factors of $\pi$ and $K$ mesons,''
  J.\ Phys.\ G {\bf 22} (1996) 1585
  [hep-ph/9508248].

\bibitem{kappa}
S.~B.~Gerasimov,
``Electroweak moments of baryons and hidden strangeness of the nucleon,''
  Chin.\ J.\ Phys.\  {\bf 34} (1996) 848
  [hep-ph/9906386].

\bibitem{PDG2020}
P.~A.~Zyla \textit{et al.} [Particle Data Group],
``Review of Particle Physics,''
PTEP \textbf{2020} (2020) 083C01

\bibitem{mass-spectrum}
 S.~Godfrey and N.~Isgur,
``Mesons in a Relativized Quark Model with Chromodynamics,''
  Phys.\ Rev.\ D {\bf 32} (1985) 189.

\bibitem{f_K}
A.~F.~Krutov,
``Electroweak properties of light mesons in the relativistic model of constituent quarks,''
  Phys.\ Atom.\ Nucl.\  {\bf 60} (1997) 1305
   [Yad.\ Fiz.\  {\bf 60} (1997) 1442].

\bibitem{Schlumpf}
 F.~Schlumpf,
``Charge form-factors of pseudoscalar mesons,''
  Phys.\ Rev.\ D {\bf 50} (1994) 6895
  [hep-ph/9406267].

\bibitem{Lattice2019}
C.~T.~H.~Davies \textit{et al.} [HPQCD],
``Meson Electromagnetic Form Factors from Lattice QCD,''
PoS \textbf{LATTICE2018} (2018) 298
[arXiv:1902.03808 [hep-lat]].

\bibitem{Roberts2017}
F.~Gao, L.~Chang, Y.~X.~Liu, C.~D.~Roberts and P.~C.~Tandy,
``Exposing strangeness: projections for kaon electromagnetic form factors,''
Phys. Rev. D \textbf{96} (2017) 034024
[arXiv:1703.04875 [nucl-th]].

\bibitem{AmendoliaK}
S.~R.~Amendolia {\it et al.},
 ``A Measurement of the Kaon Charge Radius,''
  Phys.\ Lett.\ B {\bf 178} (1986) 435.

\bibitem{Carmignotto}
M.~Carmignotto
\textit{et al.},
``Separated Kaon
Electroproduction Cross Section and the Kaon Form Factor from 6 GeV JLab
Data,''
Phys.\ Rev.\ C \textbf{97} (2018) 025204
[arXiv:1801.01536 [nucl-ex]].

\bibitem{1}
J.~He, B.~Julia-Diaz and Y.~b.~Dong,
  ``Electroweak properties of the $\pi$, $K$ and $K^*(892)$ in the three
forms of relativistic kinematics,'' Eur.\ Phys.\ J.\ A {\bf 24} (2005) 411
  [hep-ph/0503294].

\bibitem{2}
P.~Maris and P.~C.~Tandy,
  ``The $\pi$, $K^+$, and $K^0$ electromagnetic form-factors,''
  Phys.\ Rev.\ C {\bf 62} (2000) 055204
  [nucl-th/0005015].

\bibitem{4}
 Y.~Ninomiya, W.~Bentz and I.~C.~Cloet,
  ``Dressed Quark Mass Dependence of Pion and Kaon Form Factors,''
  Phys.\ Rev.\ C {\bf 91} (2015)  025202
  [arXiv:1406.7212 [nucl-th]].

\bibitem{5}
P.~T.~P.~Hutauruk, I.~C.~Cloet and A.~W.~Thomas,
  ``Flavor dependence of the pion and kaon form factors and parton distribution functions,''
  Phys.\ Rev.\ C {\bf 94} (2016)  035201
  [arXiv:1604.02853 [nucl-th]].

\bibitem{6}
C.~J.~Burden, C.~D.~Roberts and M.~J.~Thomson,
  ``Electromagnetic form-factors of charged and neutral kaons,''
  Phys.\ Lett.\ B {\bf 371} (1996) 163
  [nucl-th/9511012].

\bibitem{7}
F.~Cardarelli, I.~L.~Grach, I.~M.~Narodetsky, E.~Pace, G.~Salme and S.~Simula,
 ``Charge form-factor of pi and K mesons,''
  Phys.\ Rev.\ D {\bf 53} (1996) 6682
  [nucl-th/9507038].

\bibitem{8}
 E.~O.~da Silva, J.~P.~B.~C.~de Melo, B.~El-Bennich and V.~S.~Filho,
``Pion and kaon elastic form factors in a refined light-front model,''
  Phys.\ Rev.\ C {\bf 86} (2012) 038202
  [arXiv:1206.4721 [nucl-th]].

\bibitem{Galkin}
D.~Ebert, R.~N.~Faustov and V.~O.~Galkin,
``Masses and electroweak properties of light mesons in the relativistic quark model,''
  Eur.\ Phys.\ J.\ C {\bf 47} (2006) 745
  [hep-ph/0511029].

\bibitem{AbidinHutauruk2019}
Z.~Abidin and P.~T.~P.~Hutauruk,
``Kaon form factor in holographic QCD,''
Phys. Rev. D \textbf{100} (2019) 054026
[arXiv:1905.08953 [hep-ph]].

\bibitem{lattice}
 S.~Aoki {\it et al.} [JLQCD Collaboration],
``Light meson electromagnetic form factors from three-flavor lattice QCD with exact chiral symmetry,''
  Phys.\ Rev.\ D {\bf 93} (2016)   034504
  [arXiv:1510.06470 [hep-lat]].

\bibitem{Bijnens}
J.~Bijnens and P.~Talavera,
``Pion and kaon electromagnetic form-factors,''
  JHEP {\bf 0203} (2002) 046
  [hep-ph/0203049].

\bibitem{1805.08911}
M.~Ahmady, C.~Mondal and R.~Sandapen,
``Dynamical spin effects in the holographic light-front wavefunctions of
light pseudoscalar mesons,''
Phys. Rev. D \textbf{98} (2018) 034010
[arXiv:1805.08911 [hep-ph]].

\bibitem{JLABupgrade}
J.~Dudek {\it et al.},
``Physics Opportunities with the 12 GeV Upgrade at Jefferson Lab,''
  Eur.\ Phys.\ J.\ A {\bf 48} (2012) 187
  [arXiv:1208.1244 [hep-ex]].

\bibitem{HornRoberts2016}
  T.~Horn and C.~D.~Roberts,
``The pion: an enigma within the Standard Model,''
  J.\ Phys.\ G {\bf 43} (2016)  073001
  [arXiv:1602.04016 [nucl-th]].

\bibitem{EIC2021}
J.~Arrington
\textit{et al.},
``Revealing the structure of light pseudoscalar mesons at the Electron-Ion
Collider,'' [arXiv:2102.11788 [nucl-ex]].

\bibitem{Kiss}
L.~S.~Kisslinger, H.~M.~Choi and C.~R.~Ji,
``Pion form-factor and quark mass evolution in a light front
Bethe-Salpeter model,'' Phys. Rev. D \textbf{63} (2001) 113005
[arXiv:hep-ph/0101053 [hep-ph]].

\bibitem{QuarkCounting1}
V.~A.~Matveev, R.~M.~Muradyan and A.~N.~Tavkhelidze,
``Automodelity in strong interactions,''
Lett. Nuovo Cim. \textbf{5S2} (1972) 907

\bibitem{QuarkCounting2}
S.~J.~Brodsky and G.~R.~Farrar,
``Scaling Laws at Large Transverse Momentum,''
Phys. Rev. Lett. \textbf{31} (1973) 1153

\bibitem{Doff:mass-switching}
A.~Doff, F.~A.~Machado and A.~A.~Natale,
``Chiral symmetry breaking in QCD-like gauge theories with a confining propagator and dynamical gauge boson mass generation,''
Annals Phys. \textbf{327} (2012) 1030
[arXiv:1106.2860 [hep-ph]].

\bibitem{FarrarK}
G.~P.~Lepage and S.~J.~Brodsky,
Phys. Rev. D \textbf{22} (1980) 2157

\bibitem{JacobKiss}
O.~C.~Jacob and L.~S.~Kisslinger,
``Applicability of Asymptotic \{QCD\} for Exclusive Processes,''
Phys. Rev. Lett. \textbf{56} (1986) 225

\bibitem{Troitsky:transition}
S.~V.~Troitsky and V.~E.~Troitsky,
``Constraining scenarios of the soft/hard transition for the pion electromagnetic form factor with expected data of 12-GeV Jefferson Lab experiments and of the Electron-Ion Collider,''
Phys. Rev. D \textbf{91} (2015) 033008
[arXiv:1501.02712 [hep-ph]].


\bibitem{compDMH}
Y.~Wu, T.~Ma, B.~Zhang and G.~Cacciapaglia,
``Composite Dark Matter and Higgs,''
JHEP \textbf{11} (2017) 058
[arXiv:1703.06903 [hep-ph]].

\bibitem{compDM-F}
R.~Foadi, M.~T.~Frandsen and F.~Sannino,
``Technicolor Dark Matter,''
Phys. Rev. D \textbf{80} (2009) 037702
[arXiv:0812.3406 [hep-ph]].

\end{thebibliography}
\end{document}